\documentclass[12pt,a4paper]{article}
\usepackage{epsf}
\begin{document}
\title{On the Entropy and the Density Matrix of Cosmological Perturbations}
\author{\it I.L. Egusquiza, 
        \it A. Feinstein, \\
        \it M.A. P\'erez Sebasti\'an  and 
        \it M.A. Valle Basagoiti  \\ 
        \it Dept. of Theoretical Physics, \thanks{e-mail: wtpegegi, 
        wtpfexxa, wtbpesem, wtpvabam@lg.ehu.es} \\ 
        \it University of the Basque Country, \\ 
        \it P.O. Box 644, E-48080 BILBAO \\
        \it SPAIN}
\date{\today}
\maketitle
\begin{abstract}
We look at the transition to the semiclassical behaviour and the
decoherence process for the inhomogeneous perturbations in the
inflationary universe. Two different decoherence mechanisms appear: one
dynamical, accompanied with a
negligible, if at all, entropy gain, and the other, {\sl effectively
irreversible dephasing}, due to a rapid variation in time of the 
off-diagonal density matrix elements in the post-inflationary epoch. 
We thus settle the discrepancies in the entropy content of 
perturbations evaluated by different authors. 
\end{abstract}

\section{Introduction}

The understanding of the origin of the large-scale structure in the
universe remains one of the central problems in modern cosmology. It 
is believed that structures arise because the universe in the early
epoch was not exactly homogeneous and isotropic, but must have
contained some density irregularities. 

These density irregularities or fluctuations may be classified into two major
classes. The first one of these, the {\em primordial} fluctuations, 
corresponds to those present in the universe {\em ab initio} and which may be 
thought of as classical irregularities in the initial
structure of the universe, as for example the Mixmaster universe \cite{Mis}, 
the  similarity solutions \cite{CHw}, gravitational solitary waves \cite{CVd}, 
etc.

A more exciting possibility, and the one relevant to this paper, are the
{\em dynamically induced} fluctuations, those arising inevitably in
exact FRW models due, for example, to {\em graviton} creation in a time
varying background gravitational field \cite{HuP}. These fluctuations
are ``seen" as inhomogeneous fluctuations of the geometry. 

Since the primordial fluctuations are classical in their nature no
problem arises in interpreting the density or
temperature variations. The problem with the dynamically induced quantum 
fluctuations is more involved, however, and touches one of the
most fundamental puzzles of physics, the decoherence process in the 
process of quantum to classical transition. 

In a recent series of interesting papers \cite{LPS1}-\cite{LPS3}, 
Lesgourgues, Polarski and Starobinsky (LPS in what follows) analyze 
the evolution of inhomogeneous perturbations generated from the vacuum state
during the initial accelerated expansion of the universe. 
These authors find that the quantum fluctuations
become classical with stochastic Gaussian amplitudes, and that the
decaying mode of these fluctuations becomes exponentially small towards 
the end of the inflationary period. 
Consequently, LPS suggest that 
the exponentially small decaying mode may be
discarded for all practical purposes in this stage of the evolution, 
and the initially quantum
perturbations may be considered classical. 
It seems then that there should be no (or minimal) information loss, and,
apparently no (or minimal) entropy generation in this
quantum to classical transition. 

It is usually believed that this transition occurs when some class of coarse 
graining is enforced on the system. LPS question the 
validity of different coarse graining schemes in the following sense: 
if the dynamics of the system leads in a unitary way 
to a semiclassical behaviour is it necessary to perform a coarse graining? 
Last, but not least, stands the fact that the results obtained by LPS
on the entropy content of the perturbations are in strong disagreement
with respect to the results obtained by other authors 
(\cite{HKM} and  references therein). 

The study of the entropy of quantum fields in the cosmological context 
dates from the early days of the developments in quantum field theory
in curved spacetime (see \cite{BD} for a review up to the 80's), with
further elaboration by Hu, Kandrup, and others \cite{HP}. Currently the
language used to discuss the subject is that of squeezed
states, introduced in this context by Grishchuk and Sidorov \cite{GS},
and borrowed from the quantum optics community \cite{Sch}. The squeeze
formalism was also recently used by two of us to study the problem of
quantum tunneling under the influence of a time varying force, and the
speculation was put forward that the initial entropy gained by the
universe in the ``creation from nothing" picture may be simply evaluated by
using the squeeze parameter \cite{nos}. 

The use of the squeezing formalism should not obscure the fact that,
in the standard view, the entropy growth is always due to the particular coarse
graining one chooses; yet, interestingly enough, the whole variety
of coarse graining procedures (averaging over a period of the squeeze angle 
\cite{nos}, integrating over rotation angles \cite{BMP}, 
neglecting information about the subfluctuant variable \cite{GG}, 
setting off-diagonal elements of the density matrix to zero \cite{KOZ}) leads
in the large squeezing limit  to the same expression for the entropy generated
per mode, $S_k\approx 2 r_k$. LPS \cite{LPS3}, however, predict almost no
entropy generation for  similar systems. 

One of the main purposes of this paper is to show, within the example considered by 
LPS, that as long as the Wigner function is used to represent the evolution of
the system, as they do, the dynamical limit (certain parameter $\xi\to\infty$ in our
parametrization) and the semiclassical limit, $\hbar\to 0$, are identical, and
probably for most practical needs the system they consider may be labeled as
{\em ``dynamically semiclassical"}. However, the delicate question of the entropy
content is related to the limiting behaviour of some elements of the density matrix associated
with the system. We show that for the density matrix the
$\xi\to\infty$ and $\hbar\to 0$ limits are different, and that the
``true" decoherence is achieved only in $\hbar\to 0$ limit, due to effective
dephasing or rapid oscillations of the off-diagonal elements of the density
matrix. This we do in the following Section.

In Section 3 we consider a model universe which starts
inflating and then passes to radiation and matter dominated epochs.
Evaluating the expression for the entropy based on the density matrix
calculations we find that, due to an effective dephasing, the entropy
generated per mode is given by the usual expression $S_k\approx 2r_k$. 

Section 4 is dedicated to some final remarks. 

\section{Wigner Function and Density Matrix}

The study of the cosmological perturbations generated after the 
amplification of the vacuum fluctuations in the early stages of the
universe can be reduced to the analysis of the
evolution of a scalar field in a FRW background \cite{MFB}. 
If at some initial conformal time $\eta_{0}$ 
the field is in the vacuum state, the dynamical evolution under 
the influence of external gravitational field drives the scalar field
into excited energetic states with opposite momenta. 

One way to analyze the dynamical evolution of the cosmological 
perturbations and their quantum-to-classical transition  
is through the Wigner function formalism. The
Wigner function for a one-dimensional quantum mechanical system 
is defined by \cite{Wig} 
\begin{equation}
f_W(q,p) = \frac{1}{2\pi\hbar}\, \int_{-\infty}^{+\infty} dx\, 
\left\langle q-x/2\left|\rho \right|q+x/2\right\rangle \, e^{ip x/\hbar },
\end{equation}
where $\rho$ is the density matrix of a pure or mixed quantum state. 

The Wigner function corresponding to the evolution of the vacuum state
of a scalar field in a FRW background was evaluated by Polarski and 
Starobinsky \cite{LPS1}, and is given by
\begin{equation}
W({\bf k},-{\bf k}) = \frac{1}{\pi^2\hbar^2} \, 
\exp\left\{-\frac{|y({\bf k})|^2}{|f_k|^2}\right\}\, 
\exp\left\{-\frac{|f_k|^2}{\hbar^2}\, |p({\bf k}) - \frac{F(k)}{|f_k|^2}
y({\bf k})|^2\right\}, 
\label{fwlps}
\end{equation}
where $y({\bf k})$ is the Fourier transform of the rescaled scalar field
($y\equiv a\phi$) and $p({\bf k})$ is its canonical conjugate momentum. 
$F(k)$ and $|f_k|$ are two time-dependent parameters related to the 
variances of the field and the momentum. They are related to the 
squeeze parameter $r_k$ and the squeeze angle $\varphi_k$, the 
characteristic parameters of the Schr\"odinger picture, through \cite{LPS1}
\begin{equation}
|f_k|^2 = \frac{1}{2k}(\cosh 2r_k \,+\, \cos 2\varphi_k\,\sinh 2r_k)
\qquad
F(k) = \frac{1}{2} \sin 2\varphi_k \, \sinh 2r_k.
\end{equation}

If we decompose the field $y({\bf k})$ and the canonical conjugate
momentum $p({\bf k})$ into their real and imaginary parts, 
the two dimensional Wigner function may be further expressed 
as a product of two identical one-dimensional Wigner functions
\begin{equation}
W({\bf k},-{\bf k}) = W_1({\bf k},-{\bf k}) W_2({\bf k},-{\bf k}),
\end{equation}
where $W_1$ and $W_2$ are related to the real and imaginary part 
of the field and momentum. 

Both $W_1$ and $W_2$ may be thought of as a particular case of 
the general parametrization of the Gaussian
Wigner function (see for example Cooper et al. \cite{CHKM}):
\begin{equation}
f_W (x,p) = \frac{1}{\pi\hbar}\, \exp\left\{ 
   - \frac{x^2}{2\xi^2}\, 
   - \frac{2 \xi^2}{\hbar^2}\, \left(p-\frac{\mu}{\xi}\,x\right)^2\,
   \right\}. 
\label{fwmott}
\end{equation}
{}From now on we will use the expression (\ref{fwmott}) for convenience.

Here, comparing with Eq. (\ref{fwlps}), the parameters $\xi \equiv \sqrt{2} |f_k|$ and
$\mu \equiv F(k)/(\sqrt{2}|f_k|)$ are related to the variances of the perturbations
of the field and initially correspond to a coherent state. 
Note, as well, that this Wigner function is identical to the one obtained in 
the problem of an upside-down harmonic oscillator \cite{GP}. 

The Wigner function, in fact, is only another (particularly adequate) way of
writing all the elements of the density matrix. The relation is given by 
the following Fourier transform
\begin{equation}
\left\langle x^\prime \left|\rho \right| x  \right\rangle = \int_{-
\infty}^{+\infty} dp \, e^{ip( x^\prime - x )/\hbar} \, f_W(\frac{ 
x^\prime + x }{2},p). 
\label{rofourier}
\end{equation}
Using the expression (\ref{fwmott}), we obtain for the density matrix 
the following expression (cf. Cooper et al. \cite{CHKM})
\begin{equation}
\langle x'|\rho(\mu,\xi)|x\rangle = 
   \frac{1}{(2\pi\xi^2)^{1/2}}\, 
   \exp\left\{ -\frac{1}{4\xi^2} (x^{\prime 2} + x^2) 
               +\frac{i\mu}{2\hbar\xi} (x^{\prime 2}-x^2)
       \right\}. 
\label{rototal}
\end{equation}
It is clear that this density matrix corresponds to a pure quantum state, for
it satisfies $\rho^2 = \rho$. 

Let us analyze the behaviour of the Wigner function in two different 
limits. One, which we will call, following \cite{LPS1}, the dynamical limit and
the other the semiclassical one to be defined below. 

Since we are interested in analyzing the dynamical evolution of 
cosmological perturbations, we consider a limit corresponding 
to $\xi\to\infty$ along with $\mu/\xi$ being kept constant. This 
limit is related to the large squeezing limit characterized 
by the behaviour of the parameters $F(k)$ and $|f_k|$ which become 
unbounded. Thus, 
the dynamical limit obtained from the Wigner function (\ref{fwmott}) is
\begin{equation}
f_W^d(x,p) = \frac{1}{\sqrt{2\pi\xi^2}}\,e^{-x^2/2\xi^2}\,
\delta(p-\frac{\mu}{\xi}x).
\label{fwdyn}
\end{equation}

On the other hand, we can express the semiclassical limit in the usual way 
by analyzing the small $\hbar$ expansion in the Wigner function 
(\ref{fwmott}). It follows that
\begin{equation}
f_W^{sc}(x,p) = f_W^{d}(x,p).
\label{fwsc}
\end{equation}
Both limits represent a classical probability distribution in the 
phase space with $x$ obeying a Gaussian law,  
whereas $p$ is fixed by the value of $x$ at any instant. 

Since both limits give the same expression for the Wigner function, it
seems that the dynamical limit and the semiclassical limit are
equivalent. Thus, for any practical purpose (physical measurement of
different magnitudes) one expects to obtain the same result.  

Yet, these two limits are not completely equivalent. If one tries, say,  
to reconstruct the elements of the density matrix using
the equation (\ref{rofourier}),  one then immediately runs into 
trouble with the expression for $f_W^{sc}(x,p)$, since it is valid only 
for small $\hbar$, and, therefore, one can not use the Fourier 
transform involving $e^{ip(x'-x)/\hbar}$, given that it is not 
perturbative in $\hbar$. On the other hand, this Fourier transform 
is perfectly defined for the expression $f_W^d(x,p)$ which was 
obtained for large $\xi$ instead.

Let us reinforce the statement by looking at the behaviour of the system following 
these two limits but starting directly from the density matrix (\ref{rototal}).

The dynamical density matrix $\rho_d$, i.e., the dynamical limit of 
the density matrix as $\xi\to\infty$, is precisely the one given by 
the expression (\ref{rototal}), in the sense that $\rho_d=\rho$, 
even for very large $\xi$. The unitarity of the evolution of 
the system is obvious in this limit and $\rho_d$ satisfies ${\rm Tr}\rho_d^2 =1$. 

The leading term of the semiclassical expansion ($\hbar\to0$) of the density
matrix corresponds to
\begin{equation}
  \langle x^\prime |\rho_{sc}|x\rangle \sim 
  \left|\frac{2\xi\hbar}{\mu}\right|\, \delta(x)\delta(x^\prime).
  \label{rosc}
\end{equation}
We can see that the semiclassical limit and the dynamical limit for the density
matrix are quite different: $\rho_d \neq \rho_{sc}$, with the difference 
between them being in subleading terms in $\hbar$. 

One may still define yet a different limiting density matrix obtained from the
dynamical limit of the Wigner function $f_W^d(x,p)$, since it is quite 
``legitimate" as explained above. Performing the Fourier transform 
(\ref{rofourier}) we get
\begin{eqnarray}
\langle x^\prime |\tilde\rho_d|x\rangle & = & \frac{1}{(2\pi\xi^2)^{1/2}}\, 
   \exp\left\{ -\frac{(x^\prime + x)^2}{8\xi^2} 
               +\frac{i\mu}{2\hbar\xi} (x^{\prime 2}-x^2)
       \right\} = \nonumber\\
 & = & \langle x^\prime |\rho_d|x\rangle \exp\left\{ -\frac{(x^\prime -
x)^2}{8\xi^2}\right\} .
\label{rotilde}
\end{eqnarray}
We see that the non-diagonal elements of the density matrix obtained by Fourier
transforming the dynamical limit of the Wigner function $f_W^d(x,p)$  differ
from  those of  the pure dynamical density matrix $\rho_d$ in subleading terms. 
Furthermore, the density matrix (\ref{rotilde}) does not satisfy 
$\tilde\rho_d^2=\tilde\rho_d$ and gives ${\rm Tr}\tilde\rho_d^2 =
\infty$. This entails that $\tilde\rho_d$ neither represents a pure state, 
nor corresponds to a quantum density matrix.

We now look at the entropy. The quantum evolution of the system
corresponding to $\rho_d$ is unitary (it is governed by a quadratic
time-dependent Hamiltonian). Therefore, there is no entropy change associated
with the quantum evolution of the system in time, and since the initial state
is pure, it stays pure forever giving zero von Neumann entropy 
\begin{equation}
S_d= -{\rm Tr}\rho_d\ln\rho_d = 0.
\end{equation}

Let us now turn to the density matrix $\tilde\rho_d$ obtained from the dynamical
limit of the Wigner function. 
Since $\tilde\rho_d$ differs from $\rho_d$ in subleading off-diagonal terms,
the ``entropy" that were to be associated with it should tell us how much
information would be lost by discarding those subleading terms, or, in other
words, how strongly ``mixed"  the state given by the density matrix
$\tilde\rho_d$ is, as compared to the pure state given by
$\rho_d$.

However, the von Neumann entropy $-{\rm Tr}\tilde\rho_d\ln\tilde\rho_d$ is not well
defined here. The fact that
${\rm Tr}\tilde\rho_d^2=\infty$ and that ${\rm Tr}\tilde\rho_d=1$ implies that the matrix
has negative eigenvalues. It follows that different branches of the logarithm
would give different imaginary contributions to the von Neumann entropy, if
defined in the usual manner. We tried the following regularization: 
\begin{equation}
\tilde S_d = - {\rm Tr} \frac{1}{2} \tilde\rho_d\ln\tilde\rho_d^2. 
\end{equation}

Intuitively, one might think that the quantity just defined is
the entropy gain due to a dynamical decoherence. If so, it must be consistent 
with the calculations of LPS. However, one may easily show that this entropy 
has the wrong sign and is clearly pathological. 
The telling point here is that the state described by $\tilde\rho_d$ 
is ill-defined, and the calculations produce pathological results. 
This shows that one must be extremely careful when such a delicate
quantity as the entropy is considered. 

The dynamical limit of the density matrix $\rho_d$ given by  
expression (\ref{rototal}) is the one which correctly describes the
quantum state of the system. In fact, the system remains quantum
forever. To obtain a classical behaviour one should in principle discard 
the quantum 
interference effects. These effects are associated with those
terms of the density matrix which oscillate rapidly due 
to the smallness of $\hbar$. Thus, by averaging 
the quantum $\rho_d$ matrix over classical observation times (which is 
feasible in the cosmological context), the off-diagonal terms in 
the position representation are washed away. When the same procedure 
is applied to $\tilde\rho_d$, the density matrix obtained from the 
dynamical limit of the Wigner function (\ref{fwdyn}), we get exactly 
the same averaged density matrix, since both density matrices 
have identical rapidly oscillating phases given by the expression 
$\exp{\left\{\frac{i\mu}{2\hbar\xi} (x^{\prime 2}-x^2)\right\}}$. 

Performing the average we have just mentioned, one readily finds an 
effective coarse grained entropy 
\begin{equation}
S \approx \ln \xi, 
\end{equation}
or as expressed in  terms of the squeezing parameter $r$, 
($\xi \approx \cosh 2r$), the entropy per mode is
\begin{equation}
S \approx 2 r\,,
\label{S=2r} 
\end{equation}
in the large $r$ limit. 

This result is the same as the one we obtain by evaluating the Boltzmann 
entropy from the probability in the phase space described by the 
Wigner function $f^d_W(x,p)$ after averaging over the momentum 
$f^d_W(x)\equiv\int_{-\infty}^\infty f^d_W(x,p) dp$. The above coarse 
graining scheme is thus equivalent to tracing out the non-diagonal elements of 
the density matrix and keeping the diagonal elements alone. 

In order to clarify the issues raised by the ``bad'' behaviour of the 
elements of the matrix density in the position representation, we 
shall work in another representation which will prove to be more 
convenient.

\section{Dynamical Evolution of the Density Matrix}

As previously pointed out, we are interested in the way quantum interference effects can 
be suppressed following the dynamical evolution of the system. 
The behaviour of the system at large times, i.e., when the 
inflationary epoch is followed by radiation and matter dominated periods, 
is similar to that of a parametric oscillator whose time-dependent potential 
is slowly turned off. Thus, it seems that the most adequate basis 
to study the evolution would be that of the harmonic oscillator. 

In the Schr\"odinger representation the dynamics of the state is
given by the time evolution operator $|0,\eta\rangle_S \equiv
S|0,\eta_0\rangle$. The time operator $S$ is the
usual two-mode squeeze operator and the time evolution of the density matrix 
expressed in the harmonic oscillator basis is given by \cite{Sch, CHKM}
\[
\langle n=2l|\rho|m=2l^\prime\rangle = \langle n|S(r,\varphi)|0\rangle
                          \langle 0|S^\dagger (r,\varphi)|m\rangle 
\]
\begin{equation}
= (-1)^{l+l^\prime} \, 
    \frac{((2l)!\, (2l^\prime)!)^{1/2}}{2^{l+l^\prime}\, l!\,l^\prime !} \, 
    \frac{(\tanh r)^{l+l^\prime}}{\cosh r}\,   e^{2i\varphi(l-l^\prime)}, 
\label{rhonm}
\end{equation}
for each mode $k$. For convenience we have dropped the sub-index $k$, 
and $r$ and $\varphi$ are the usual squeeze parameter and squeeze 
angle, respectively. 
The only non vanishing diagonal elements are 
\begin{equation}
\langle 2l|\rho|2l\rangle = \frac{(2l-1)!!}{2^l\, l!} \, 
\frac{(\tanh r)^{2l}}{\cosh r}.
\label{diag}
\end{equation}
Note that only the non-diagonal elements of the density matrix 
depend on the squeeze angle $\varphi$.

The time dependence of the density matrix elements in this basis is given by 
the evolution of the squeezing parameters which can be obtained through 
the interrelation between the Heisenberg and Schr\"odinger pictures \cite{Sch, AFJP}. 
The equations of motion for $r$, $\varphi$ and $\theta$ are \cite{LPS1, Sch, AFJP}
\begin{eqnarray}
r^\prime &=& \frac{a^\prime}{a} \, \cos 2\varphi,\nonumber \\
\varphi^\prime &=& -k\, -\frac{a^\prime}{a} \, \coth 2r\, \sin 2\varphi, 
\label{eqsmov}\\
\theta^\prime &=& +k\, +\frac{a^\prime}{a} \, \tanh r\, \sin 2\varphi.\nonumber
\end{eqnarray}
The squeeze parameter $r$ grows indefinitely towards the end of the
inflationary period while the squeeze angle $\varphi$ approaches a
constant value. Thus, the non-diagonal elements of $\rho_d$ do not
oscillate, and no dephasing takes place, as was correctly pointed out by 
LPS. 

However, let us now push the evolution of the system forward, pass the
inflationary period by matching it to a radiation dominated one 
followed by a matter dominated expansion as, for example, was considered 
by Grishchuk and Sidorov \cite{GS}. It follows then that the 
behaviour of the squeeze parameter $r$ and the squeeze angle $\varphi$ 
are qualitatively different from the one we see 
during the end of the inflationary period. 

To be specific, we use Grishchuk's and Sidorov's \cite{GS} simple 
model of the universe containing the three stages of expansion just 
described above 
\begin{eqnarray}
a_{i} & = & -\frac{1}{H\eta} \quad\qquad\qquad\qquad\qquad\quad\quad 
                              (-\infty<\eta\le\eta_1<0), \nonumber \\
a_{r} & = & \frac{1}{H\eta_1^2} \, (\eta - 2\eta_1) \;\quad\quad\qquad\qquad\qquad 
                               (\eta_1\le\eta\le\eta_2), \label{afactor} \\
a_{m} & = & \frac{1}{4H\eta_1^2 \, (\eta_2-2\eta_1)} \, (\eta - 4\eta_1 + \eta_2)^2
                             \quad (\eta_2\le\eta<\infty), \nonumber
\end{eqnarray}
where $a_{i}$, $a_{r}$ and $a_{m}$ represent the scale factor during the
inflationary era, radiation dominated period and matter dominated period
respectively, and $\eta_1$ and $\eta_2$ are times where the
transition between two consecutive periods takes place. 

We have solved numerically  equations (\ref{eqsmov}) under  
conditions (\ref{afactor}) and found the following behaviour of the
squeeze parameters: the squeeze parameter $r$ approaches its
maximum value at the end of the inflationary period, oscillates around this large maximum
value during the radiation dominated period and 
finally sets to a constant in the matter dominated epoch. 
This behaviour can be seen in the Figure 1, where the 
parameter $r$ is represented as a function of conformal time $\eta$. 

	  \begin{figure}[ht]
	  \begin{center}
	  \epsfxsize=6	cm
	  \epsfbox{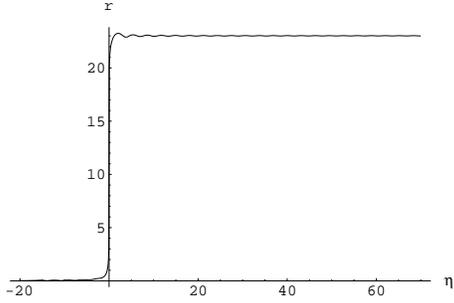}
	  \caption{Dependence in time of the squeeze parameter	$r$.
                The following values for the
                parameters appearing in the dynamical equations are chosen: 
                $\eta_0=-20$, $\eta_1=-10^{-5}$, $\eta_2=20$ and the wave
                number has been set to $k=1$.}
	  \end{center}
	  \end{figure}

On the other hand, as is apparent from  Eq. (\ref{rhonm}), the 
squeeze angle $\varphi$ shows itself as an angular phase 
$e^{2i\varphi(l-l^\prime)}$ in the non-diagonal elements of 
the density matrix. This angular 
parameter $\varphi$ grows linearly with time as 
$\varphi \approx \varphi_0 - k \eta$ leading to oscillating behaviour of 
the off-diagonal density matrix elements. Thus, the dynamical evolution 
of the model suggests a natural way to average these elements to zero. 
It is clear then,  that sooner or
later an effective dephasing process will take place. A typical behaviour of 
the phases of the off-diagonal density matrix elements is presented in Figure 2.

	  \begin{figure}[ht]
	  {\noindent
	  \begin{minipage}[t]{.5\linewidth}
	  \begin{center}
	  \epsfxsize=6cm
	  \epsfbox{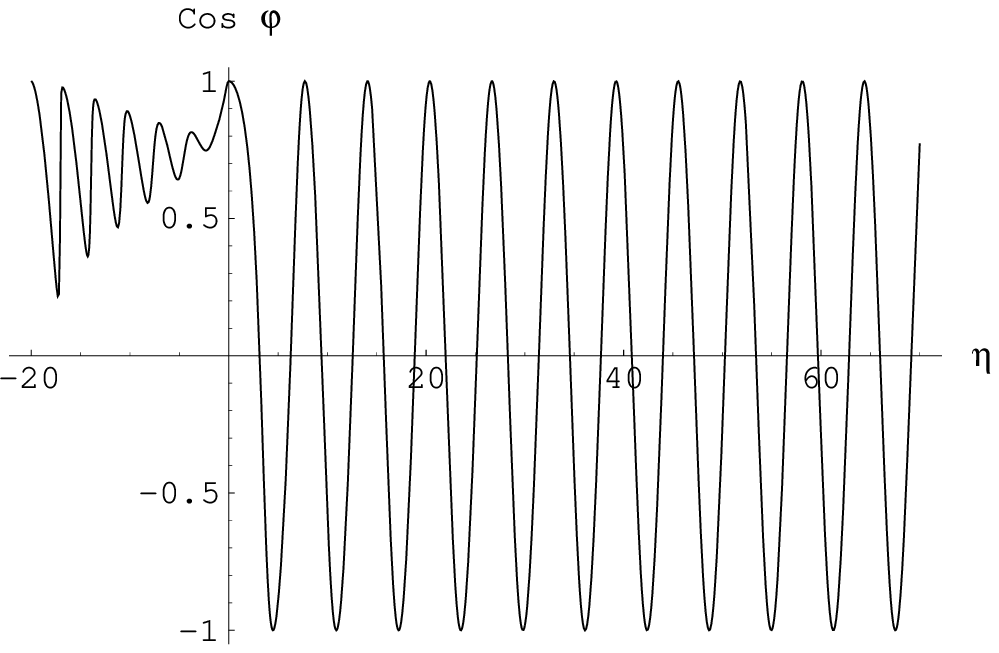}
	  \end{center}
	  \end{minipage}\hfill
	  \begin{minipage}[t]{.5\linewidth}
	  \begin{center}
	  \epsfxsize=6cm
	  \epsfbox{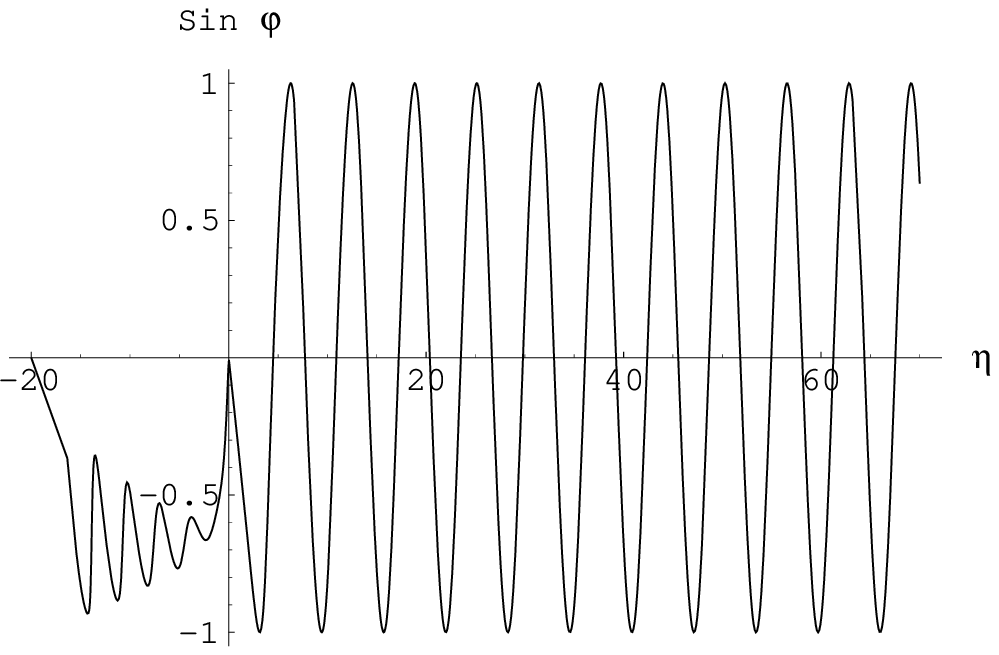}
	  \end{center}
	  \end{minipage}}
	  {\noindent
	  \begin{minipage}[t]{.5\linewidth}
	  \begin{center}
	  Fig.	 2a
	  \end{center}
	  \end{minipage}\hfill
	  \begin{minipage}[t]{.5\linewidth}
	  \begin{center}
	  Fig.	 2b
	  \end{center}
	  \end{minipage}}
	  \caption{The	figure 2a represents the real part of the phase
                   of the off-diagonal density matrix elements.
                   Imaginary part of the phase is represented on figure 2b.}
	  \end{figure}

As a consequence of the above dephasing process one may proceed in
calculating the entropy in the usual way by dropping the non-diagonal
elements of the density matrix. 
\begin{eqnarray}
S &=&  -\sum_{l=0}^\infty \langle 2l|\rho|2l\rangle \, 
                        \langle 2l|\ln\rho|2l\rangle \nonumber \\  
   &=& -\sum_{l=0}^\infty \frac{(2l-1)!!}{2^l\, l!}\, 
                                      \frac{(\tanh r)^{2l}}{\cosh r}\, 
                    \ln\left[ \frac{(2l-1)!!}{2^l\, l!}\, 
                              \frac{(\tanh r)^{2l}}{\cosh r}\right].
\end{eqnarray}
For convenience, the latter expression can be separated into two pieces
\begin{equation}
S_1 = -\sum_{l=0}^\infty \frac{(2l-1)!!}{2^l\, l!}\, 
                                      \frac{(\tanh r)^{2l}}{\cosh r}\, 
                      \ln\left[\frac{(\tanh r)^{2l}}{\cosh r}\right], 
    \label{s1} 
\end{equation}
and
\begin{equation}
S_2 = -\sum_{l=0}^\infty \frac{(2l-1)!!}{2^l\, l!}\, 
                                      \frac{(\tanh r)^{2l}}{\cosh r}\, 
                      \ln\left[\frac{(2l-1)!!}{2^l\, l!}\right], 
    \label{s2}
\end{equation}
where the first one readily gives
\begin{equation}
S_1 = \ln(\cosh r) \, - \, \sinh^2(r) \,\ln(\tanh r).  
\end{equation}
To evaluate the second term we expand expression (\ref{s2}) in terms of
gamma functions and get for $r\gg 1$ 
\begin{equation}
S_2 = - \frac{1}{\cosh r} \, \sum_{l=1}^\infty 
        \frac{\Gamma(2l)}{2^{2l-1} \, l\, (\Gamma(l))^2} \, (\tanh r)^{2l}\,
        \ln\left[\frac{\Gamma(2l)}{2^{2l-1} \, l\,
        (\Gamma(l))^2}\right]. 
\label{s_2}
\end{equation}
With the aid of the Stirling's formula 
$
\ln\left[\frac{\Gamma(2l)}{2^{2l-1} \, l\, (\Gamma(l))^2}\right] \sim 
      -\frac{1}{2}\, \ln (l\pi), 
$
we obtain
\begin{equation}
S_2 \sim  \frac{1}{2\cosh r} \, \sum_{l=1}^\infty 
        \frac{(\tanh r)^{2l}}{\sqrt{l\pi}} \, 
        \ln(l\pi). 
\end{equation}
Using the McLaurin asymptotic method \cite{asymp} for large $r$ we get 
\begin{eqnarray}
S_2 &\sim& - \frac{1}{2\sqrt{2}} \, \frac{1}{\cosh r\, \sqrt{-\ln\tanh r}}\,
             \left( \gamma + \log(\frac{-8\ln\tanh r}{\pi}) \right) + C +\nonumber\\     
     &+& O[-\ln\tanh r],
\end{eqnarray}
where $C$ is a constant.  

Finally, the total coarse grained entropy is given by 
\begin{eqnarray}
S &\sim & \; \ln\cosh r - \sinh^2 r \ln\tanh r+\,C\,- \nonumber\\ 
  &-& \frac{1}{2\sqrt{2}} \, \frac{(-\ln\tanh r)^{-1/2}}{\cosh r}\,
      \left( \gamma + \log(\frac{-8\ln\tanh r}{\pi}) \right), 
\end{eqnarray}
and in the limit $r\to\infty$ we  obtain 
\begin{equation}
S \approx 2r.
\end{equation}

\section{Conclusions}

Before closing we feel that some final remarks are in order. 

The first
lesson learnt from our analysis is that one must be careful with the
equivalent expressions obtained within different limiting procedures.
From the example with the Wigner function we have seen that although the
two expressions coincide (the one obtained in the dynamical limit, $\xi\to\infty$, of
the Wigner function and the other in the semiclassical one, $\hbar\to 0$), 
one cannot use the last one to reconstruct the elements of
the density matrix. The reason is simple: the reconstruction of
the density matrix involves again the parameter $\hbar$. For any other
calculation, these two limits may be perfectly equivalent, but if one is
interested in the entropy of the system, the procedure fails! 
One may easily see this also by
evaluating the two limits starting directly from the expression for the
density matrix elements (\ref{rototal}). The difference between the
limits is evident from expression (\ref{rosc}).

Going somewhat further, we were interested in seeing whether, by
comparing the elements of the density matrix of
the pure state taken in the late time limit and those ``legitimately" 
obtained from the dynamical limit
of the Wigner function, we can get a measure of the information loss
due to dynamical decoherence. Intuitively, this measure should have 
been consistent with the small gain in the entropy due to the phase-space 
volume increase as described by LPS \cite{LPS3}. Unfortunately the entropy 
expression calculated in this way is ill-defined. 

To evaluate properly the entropy generation for each mode of the
perturbation we study the time evolution of the density matrix in the
harmonic oscillator basis. Physically, this seems to be the correct
basis since at late times, during the post-inflationary behaviour, the
time dependent potential governing the dynamics of the modes
switches off adiabatically. We explicitly see the dephasing behaviour
due to the rapid oscillations of the off-diagonal elements of the
density matrix, and calculate the entropy growth recovering the result
$S_k\approx 2r_k$. It is important to note, however, that each mode has its
proper time scale for dephasing depending on its wavenumber $k$, yet 
(almost) all modes, as mentioned above, sooner or later contribute to the
process. 

Decoherence is normally understood as the vanishing of the off-diagonal
density matrix elements in a particular basis and is accompanied by the
entropy growth due to the loss of information encoded in those
elements. We have evaluated this entropy and found it consistent with
previous results. 

On the other hand, we have found that the entropy associated with the
{\sl dynamical decoherence process} evaluated on the basis of the 
dynamical limit of the Wigner function is ill-defined and probably carries no 
useful information about the statistical properties of the system. 

\section*{Acknowledgements}
This work was supported in part by the Spanish government through the
CICYT under projects AEN96-1668 (I.L.E. and M.A.V.B.) and PB93-057 
(A.F.). We are grateful to David Polarski, for discussions and 
correspondence.

\end{document}